# Dark Matter, Mass Scales Sequence, and Superstructure in the Universe (with extension and summary)

Wuliang Huang

Institute of High Energy Physics, Chinese Academy of Sciences,
P.O.Box 918(3), Beijing 100049, China

Xiaodong Huang
Department of Mathematics, University of California, Los Angeles
Los Angeles, CA 90095, U.S.A.

## Abstract

The original intention of the first version of this paper is to search for the mass rang of dark matter particles (fermions or bosons) $m_d$ according to the material state equations in the evolved universe. It shows the mass rang of $m_d$ is around $10^{-1}eV$ , i.e. into H-decoupling area. The dark matter particles with low (light) mass are related to the superstructure and belong to a mass scale sequence in the universe. The cosmic structure is not only from primary perturbation of early universe but also from perturbation and Jeans density wave of dark matter in the superstructure. The environment inside superstructure (SS) speed up the formation of sub-structure in SS by H-decoupling period. If dark matter is dominated by bosons, the cosmology principle is needed; if dark matter is dominated by fermions, it tends to spherical universe (finite universe).

The second section (extension section) of this paper describes the extension of the mass scales sequence (mass tree) and describes new particles ($u$ -particle, Planck particle, $A$ -particle, $\delta$ -particle, *graviton*, $\pi'$ -particle, heavy electron…). This can be used to explore the state of matter with super-high density, inflation, lightest black hole (LBH), BSM, CGB etc. in the early universe. Measuring the rate of change of $c$ and $\hbar$ at the present time is one way to check whether the cosmological constant/dark energy exists or not.

The third section of this paper is a summary. From mass tree, the evolution of the universe is described by three stages: chaos, inflation and expansion. The first two stages have $c$ mutations and the inflation appears as a step by step fission process of black holes. The dark matter particles with low mass ($\nu$ and $\delta$ ) are described in a dual/two-fold SM with new symmetry and new interaction, and $\delta$ -particle is like inert neutrino but has baryon number (L-B conservation). We emphasize how to search for $\delta$ -particle, how to research critical energy, critical density $\rho_{cr}$ , background particles, and spherical universe. $\rho_{cr}$ relates to a type of pseudo-balance black holes/celestial bodies. Minimum black hole radius $r_{\min} = r_p$ means we live in a sphere universe, which belong to a big universe, mainly characterized by proton. It appears in a new manner for the super-high energy physics, which may determine the applicable region for the modern physics including relativity.

PACS numbers: 95.35. +d, 98.65.D, 98.80. -K

Keywords: large number, mass scales sequence (mass tree), superstructure, dark matter, critical energy, critical density, cosmological constant, lightest black hole (LBH), BSM, inflation, background particles, spherical universe

To current knowledge, the stable elementary particles existing in nature are nucleon $n$ (mass $m_n$), electron $e$, photon $\gamma$ and neutrino $\nu$ ($\nu$ can be put into the category of dark matter). Dark matter particles $d$ (mass $m_d$) could be fermions $f$ with mass $m_f$ or bosons $b$ with mass $m_b$. Because $e$ and $\gamma$ make only a small contribution to the total mass of the universe at the present time, we are confronted with a multi-component ($n+d$) universe, in which the typical mass scale is the solar mass $M_\Theta$. On the other hand, from the fundamental physical constants (except electric charge) the speed of light $c$, the gravitation constant $G$ and the Planck constant $\hbar$, a mass scale (Planck mass): $m_{pl} \sim \sqrt{\frac{\hbar c}{G}} \sim 10^{19} GeV$ can be deduced. First, we discuss the internal relations of $m_{pl}$, $m_n$, $m_d$, and $M_\Theta$.

When a star collapses into a neutron star, it can be simplified as a degenerate system composed of neutral nucleons (fermions). Inside the neutron star, the boundary momentum of fermions comes to the maximum value $m_n c$. At this time the total number of nucleons in the star with a volume of $V$ is $N = \frac{gVp_o^3}{6\pi^2\hbar^3}$ [1], where $g=2$ and $p_o = m_n c$. The neutron star mass is $M = Nm_n$, and the minimum mass of a black hole (BH) collapsed from a star is $M_{star} \sim M$. Since the classical black hole radius (CBHR) is $r_{star} \sim \frac{GM_{star}}{c^2}$ and the nucleon radius is $r_n \sim \frac{\hbar}{m_n c}$,

$$M_{star} \sim \frac{m_{pl}^3}{m_n^2} \sim 10^0 M_\Theta \qquad (1)$$

(** annotation at April, 2020: *Eq(1) can be expressed for degenerate celestial bodies as* $M_n \sim \frac{m_{pl}^3}{m_n^2}$*, for n=0,1,2,... and used to next sections)*

This is the scale of the free stream scale (FSS) of nucleons at the early era of the universe. From Eq (1), $\frac{M_{star}}{m_n} \sim (\frac{m_{pl}}{m_n})^3$, and

$$\frac{m_{pl}}{m_n} \sim \sqrt{\frac{\hbar c}{Gm_n^2}} \equiv A \qquad (2)$$

This is the large number used in this paper, $A \sim 10^{19}$, so

$$\frac{M_{star}}{m_n} \sim A^3, \ \frac{r_{star}}{r_n} \sim A \qquad (3)$$

From $c, G, \hbar$, a length scale (Planck length) $r_{pl} \sim \sqrt{\frac{\hbar G}{c^3}} \sim \frac{Gm_{pl}}{c^2}$ can be composed which also has the form of CBHR. Suppose a Planck particle has mass $m_{pl}$ and radius $r_{pl}$, then

$$\frac{m_n}{m_{pl}} \sim A^{-1},\ \frac{r_n}{r_{pl}} \sim A \tag{4}$$

That is to say, a nucleon can contain $\sim 10^{57}$Planck particles (string phenomenology)[2] as a star can contain $\sim 10^{57}$nucleons, but $m_n << m_{pl}$. Why is $m_n << m_{pl}$? With the aid of the large number $A$, the nucleon radius can be expressed in a CBHR form, $r_n \sim \frac{\tilde{G}m_n}{c^2}$, where $\frac{\tilde{G}}{G} = A^2 \sim 10^{38}$. This is just the right ratio of two nucleons' strong interaction force to their gravitation interaction force. So, a nucleon is like a "strong BH" under a "strong gravitation" interaction with a "strong gravitation constant" $\tilde{G}$, and confines "strong signals".

The main results discussed above can be summarized as follows:

| radius | mass | CBHR | FSS |
|---|---|---|---|
| | $m_n$ | | |
| $r_{pl}$ | $m_{pl} = Am_n$ | $r_{pl} \sim \frac{Gm_{pl}}{c^2}$ | |
| $r_n = Ar_{pl}$ | | $r_n \sim \frac{\tilde{G}m_n}{c^2}$ | |
| $r_{star} = A^2 r_{pl}$ | $M_{star} = A^3 m_n$ | $r_{star} \sim \frac{GM_{star}}{c^2}$ | $\frac{m_{pl}^3}{m_n^2} \sim M_{star}$ |

From this table, one could infer that the next mass scale is $M_F = A^4 m_n \sim 10^{19} M_\Theta$, which is the superstructure scale in the universe[3]-[8]. In a ($n+d$) universe, the scale of $M_F$ may also have a connection with another FSS, $M_F \sim \frac{m_{pl}^3}{m_d^2}$. From the equations $M_F = A^4 m_n \sim 10^{19} M_\Theta$ and $M_F \sim \frac{m_{pl}^3}{m_d^2}$, we can obtain $m_d \sim A^{-0.5} m_n \sim 10^{-1} eV$. This means that the mass of non-baryonic dark matter particles (NBDMP) is in an $10^{-1}eV$ order of magnitude. Thus, there is a sequence of mass scales from micro-cosmos to macro-cosmos: $A^{-1.5}m_{pl}$, $A^{-1}m_{pl}$, $A^2 m_{pl}$, and $A^3 m_{pl}$, corresponding to the mass scales of dark matter particles, nucleons, stars and superstructures respectively.

We shall now directly calculate the mass and state of NBDMP to check the above deduction about the mass of NBDMP and the sequence of mass scales. If the NBDMP is dominant in the universe at the present time, the direct calculation can be simplified and done for a one-component universe composed of NBDMP only. First, we can suppose that the NBDMP are stable and weakly interacting massive fermions ($f$) and calculate the mass $m_f$, as well as the state parameters (chemical potential $\mu_f$ and temperature $T_f$) of $f$-particles using three equations. Under the standard cosmological model and the non-relativistic condition, the state equation is[1]

$$\rho_f = \frac{g m_f^{\frac{5}{2}} (kT_f)^{\frac{3}{2}}}{2^{\frac{1}{2}} \pi^2 \hbar^3} \int_o^\infty \frac{\sqrt{Z}dZ}{\exp(Z-\gamma)+1} = \Omega_f h^2 \rho_c \qquad (5)$$

where $g$ is the variety number of $f$-particles, $\gamma \equiv \frac{\mu_f}{kT_f}$. The critical density of the universe is $\rho_c = \frac{3H_{100}^2}{8\pi G}(1+z)^3$, $H_{100} = 100 km \cdot \sec^{-1} \cdot Mpc^{-1}$, $h = \frac{H_o}{H_{100}}$ ($z$ is red shift). The evolutionary equation of temperature is

$$T_f \cong \frac{kT_{f0}T_{\gamma 0}}{\xi m_f c^2} \cdot (1+z)^2 = \frac{k\tilde{T}_{f0}T_{\gamma 0}}{m_f c^2} \qquad (6)$$

where $T_{\gamma 0}$ is the microwave background temperature, $T_{\gamma 0} = 2.7^\circ K$. $T_{f0}$ is the $f$-particles temperature when $m_f = 0$. $\tilde{T}_{f0} = \frac{T_{f0}}{\xi} \cdot (1+z)^2$. Here, $\xi$ is a phenomenological parameter representing non-relativity, $\frac{kT_{\gamma 0}}{m_f c^2} \le \xi \le 1$. The third equation is in relation to the superstructure of the universe mentioned above. In the last decade some reports related to the very large scale structure (superstructure) in the universe were published[3]-[7]. Specifically, reports about the periodic superstructure[6],[7] enlightened us. We think the formation of such structure may be related to the gravitation and the hydrodynamic effect in cosmic medium. Since the scale of the superstructure has been 1% - 10% of the present horizon, it is appropriate to adopt the sound velocity $v_s$ in cosmic medium: $v_s = 0.01c - 0.1c$.

$$v_s = \sqrt{\frac{10}{9} \frac{kT_f}{m_f} \cdot \int_0^\infty \frac{Z^{3/2}dZ}{\exp(Z-\gamma)+1} \Big/ \int_0^\infty \frac{\sqrt{Z}dZ}{\exp(Z-\gamma)+1}} \sim 0.01c - 0.1c \qquad (7)$$

From the three equations, we can obtain the results[8]:

$$v_s = 0.01c - 0.1c$$
$$m_f = 10^{-1} - 10^{-2} eV$$
$$\mu_f = 10^{-5} - 10^{-4} eV$$
$$\xi T_f = 10^{-3} - 10^{-2} \ {}^\circ K$$
$$\gamma = 10^1 - 10^2$$

for $z = 0$ and $w \equiv \dfrac{g}{\Omega_f h^2} = 1 \sim 80$. The above values of $\gamma$ mean that the $f$-particles are in a degenerate state. Under the degenerate approximation, we have $m_f^4 = \dfrac{2\pi^2}{3^{\frac{1}{2}}} \cdot \dfrac{\hbar^3 \rho_c}{w v_s^3}$, $\mu_f = \dfrac{3}{2} m_f v_s^2$; $\dfrac{T_f}{\tilde{T}_{f0}} \cong \dfrac{kT_{\gamma 0}}{m_f c^2}$. So, the values of $m_f$ is not related to $z$, and is not sensitive to the parameters $g$, $\Omega_f$, $h$ ($m_f \propto w^{-\frac{1}{4}}$).

Because the periodic superstructures in the universe[6],[7] can be described by Jeans length $\lambda_J \sim \dfrac{v_s}{\sqrt{G\rho_f}} \sim 10^2 Mpc$ [6], the concrete value of $\dfrac{v_s}{c}$ at $z = 0$ in an equivalent homogeneous universe is: $\dfrac{v_s}{c} \sim \dfrac{\lambda_J}{r_H} \sim 0.01$, where $r_H$ is the present horizon. From the above calculation, $m_f$ is indeed $\sim 10^{-1} eV$.

However, the maximum scale of superstructures from observations is $\sim 10^3 Mpc$ [6], corresponding to a typical mass scale $M_F$ mentioned above. The superstructures appeared during H-decoupling, when $m_f \sim 10^{-1} eV$. Once a superstructure has broken away from the cosmic expansion, the inner environment is like a quasi-static universe. Thus, celestial bodies with different scales originating from various cosmic perturbations were rapidly produced in the superstructure[9]. In the formation of celestial bodies, one of the essential conditions is that the particles of cosmic medium must be in a non-relativistic state with an average thermal velocity $\bar{v} \sim v_s \sim 0.1c$. From Eq (6) and Eq (7), we know at this time $z$ is $\sim 10$, and may be near the time that superstructures broke away from the cosmic expansion. Thus, the existence of stable NBDMP with mass $\sim 10^{-1} eV$ is not in contradiction with the recent report about the existence of galaxies at large redshift $z \sim 10$ [10].

Another way to calculate the matter state of the $f$-particle is to substitute the evolutionary equation of temperature, Eq (6), with the concrete chemical potential value of the $f$-particle. According to the iso-entropic hypothesis of the evolution of the universe, the entropy per $f$-particle (S/N) is a constant:

$$S/N=\frac{k}{3}\cdot\{\int_o^\infty Z^{\frac{3}{2}}(Z+\frac{2m_f c^2}{kT_f})^{\frac{3}{2}}\frac{dZ}{\exp(Z-\gamma)+1}$$
$$+3\int_o^\infty (Z+\frac{m_f c^2}{kT_f})\sqrt{Z(Z+\frac{2m_f c^2}{kT_f})}\frac{(Z-\gamma)dZ}{\exp(Z-\gamma)+1}\}$$
$$\div\int_o^\infty \frac{dZ}{\exp(Z-\gamma)+1}\cdot(Z+\frac{m_f c^2}{kT_f})\sqrt{Z(Z+\frac{2m_f c^2}{kT_f})}$$

Under relativistic condition, S/N = $\frac{k}{3}\cdot\frac{4J_3-3J_2\cdot\gamma}{J_2}$. Under non-relativistic condition, $S/N=\frac{k}{3}\cdot\frac{5J_{3/2}-3J_{1/2}\cdot\gamma}{J_{1/2}}$, where $J_a\equiv\int_o^\infty\frac{Z^a dZ}{\exp(Z-\gamma)+1}$. On the other hand, there are typically two situations for $\mu_f$ [11] under relativistic condition: the first is $\gamma>20$ (degenerate state) and the second is $\gamma=0$. From calculations we know that for $\gamma>20$, the non-relativistic fermions created from relativistic fermions will still be close to a degenerate state. However, for $\gamma=0$, the value of non-relativistic fermions will become $\gamma=-1.62$ instead of zero when created from relativistic fermions. We can substitute such value of $\gamma$ for the evolutionary equation of temperature and obtain the approximate expressions of $m_f$, $T_f$, and $\mu_f$:

$m_f^4=(\frac{10\pi}{3})^{\frac{3}{2}}\frac{\hbar^3\rho_c}{wv_s^3\cdot\exp(\gamma)}$, $T_f=\frac{3}{5}\frac{m_f v_s^2}{k}$, $\mu_f=\frac{3\gamma}{5}m_f v_s^2$. Based on the value of $\gamma=-1.62$ and the same parameter ranges as before, the calculated values of $m_f$ and $|\mu_f|$ are approximately unchanged.

If the NBDMP are bosons, all of the approximate equations and results for $f$-particles with negative chemical potential are still suitable for $b$-particles since the chemical potential of bosons are negative. However, the subscript $f$ must be substituted by $b$, and the term $[\exp(Z-\gamma)+1]$ must be substituted by $[\exp(Z-\gamma)-1]$. When $\mu_b=0$, $b$-type dark matter particles will have a minimum mass $m_b=(4.77\frac{\hbar^3\rho_c}{wv_s^3})^{\frac{1}{4}}\sim10^{-1}eV$, which is different from the ordinary axions.

To summarize: (1) There is a category of stable NBDMP in the universe at the present time, which is related to the superstructure of the universe. These particles are fermions or bosons. In either case, we deduced that the particle mass is $\sim 10^{-1} eV$ and the absolute value of its chemical potential is $<< 10^{-1} eV$. These results are not in contradiction with the existence of galaxies at large red shift $z \sim 10$, nor with the dip phenomena of the ultra-high energy primary cosmic ray spectrum at $\sim 10^{15} eV$ ("knee") corresponding to fermion NBDMP $\nu$ and at $\sim 10^{18} eV$ ("ankle") corresponding to boson NBDMP ***or to the fourth stable elementary particle - fermion $\delta$. If dark matter is dominated by bosons, cosmological principle is necessary. If dark matter is dominated by fermions, it favors the spherical universe theory*. (2) This paper is consistent with our previous works[8],[12],[13]. If the NBDMP with mass $\sim 10^{-1} eV$ do exist in the universe, they can be used to explain the large scale stream[8] and the filament[12] in the universe, as well as the flatness of the rotational velocity distribution in spiral galaxies[13]. (3) If the $f$-particles are neutrinos, the neutrino mass is also $\sim 10^{-1} eV$ [14] since the value of $m_f$ is not sensitive to parameter $\Omega_f$. (4) If the superstructure scale $M_F$ indeed exists in the universe, the cosmology principle must be based on superstructures. Therefore, the observed value of the Hubble constant $H_o$ and the cosmological constant $\Lambda$ must take into consideration the influence of the superstructure $M_F$. That is, the value of $\Lambda$ from the data about the SNe Ia[15] could still be equal to zero[16]. (5) The concept of large number was introduced by P.A.M.Dirac[17]. In this paper the large number $A$ connects microcosms with macrocosms by a sequence of mass scales, and also contributes to probe the precise structure of the nucleon[2].

(6) Under the framework of this paper, there is no possibility for stable NBDMP with heavy mass as the dominant component of the universe at the present time. If heavy mass NBDMP exist in the halo region of our galaxy by a violent relaxation process, why would our galaxy be a spiral galaxy instead of an elliptical galaxy? If these heavy particles (perhaps SUSY particles) are not recognized in the experiments during the next decade (as the present status about 17 keV neutrinos or monopoles), the NBDMP discussed in this paper and the cold universe (***should use words "spherical universe" or "ball universe"*) will be progressively researched again[18].

---

** For annotation at April, 2020

---

Email address: huangwl39@yahoo.com, huangwl@ihep.ac.cn
xhuang@ucla.edu

# Extension

We have left the content of the original paper (the above section, i.e. arXiv: astro-ph/9909321v1) unchanged, and make extensions below for contrast.

## Starting Point

(1)The universe is evolving and the space-time background (vacuum background) is also evolving. We can arrive at an energy scale of the universe in a laboratory, but we can't produce the space-time background of that time $t_{univ}$. In a laboratory, there only exists the present space-time background.

(2) The homogeneous and isotropic cosmological principle is an approximation. The explanation of the expansion of the universe is related to this principle. Any two points in the universe do not always reflect this expansion: two points located in the same galaxy, for example. We must correctly select a set of representative points, which approximately satisfy the cosmological principle. For example, the representative point can be a SN Ia, a galaxy, a cluster of galaxies, a superstructure, etc. Different types of representative points make up different sets, and different sets reflect the same universe. We cannot be sure if from different sets we can obtain the same Hubble constant $H_0$ , the same regression parameter $q_0$, and the same cosmological constant $\Lambda$,etc.

To reflect the cosmological principle, the universe essentially appears as a set of representative points. There is no "continuity" concept in physical cosmology and no singularity at the beginning of the universe. The representative point can also be an island universe (spherical universe).

(3)The transparent process (hydrogen decoupling process) of the universe is a long and complex one. We cannot be certain if the anisotropy of CMB is from the quantum perturbations in the early universe. It can also be from the gravitational instability process around hydrogen decoupling epoch ($10^0 - 10^{-1} eV$) in a multi-components universe composed by some types of stable elementary particles with low mass.

(4) The particles in particle physics are lengthy and complex, but the stable massive particles are simple. The stable massive particles for “bright” matter is electron $e$ and proton $p$. The stable massive particles for dark matter are neutrino $\nu$ with mass $\sim 10^{-1} eV$ and delta particle $\delta$ with mass $\sim 10^{0} eV$ (Ref {1}).

(5) We know little about the states of matter, which has density greater than nucleus density $\rho_{nucl}$. In the early universe, the density of the universe $\rho_{univ}$ is much greater than $\rho_{nucl}$ ($\rho_{univ} >> \rho_{nucl}$). We assume that the Fermi and Bose statistical distributions hold and that the black body radiation spectrum is also correct. So, $\hbar^3 c^3 = const.$ and

$$\hbar c = const. \qquad \{1\}$$

(6) The speed of light is not a global physical quantity. It is dependent on local space-time background (vacuum background), $c = c(t_{univ}, \rho_{univ})$. The constancy of the speed of light is an approximation.

At the early epoch of the universe $\rho_{univ} >> \rho_{nucl}$, the speed of light actually represents the transfer speed of interactions. From Eq {1}, if $c \to \sim\infty$, we have $\hbar \to \sim 0$, then $m_{pl} = \sqrt{\frac{\hbar c}{G}} = const.$ and $r_{pl} = \sqrt{\frac{G\hbar}{c^3}} \to \sim 0$.

## The extension of mass scales sequence

From the above original version, we have a sequence of mass scales from micro-cosmos to macro-cosmos: $A^{-1.5} m_{pl}$, $A^{-1} m_{pl}$, $A^{2} m_{pl}$, and $A^{3} m_{pl}$, corresponding to the mass scales of neutrinos, protons, stars, and superstructures respectively. From Ref {1}, we have two extensions of mass scales: $M_{cr} \sim A^4 m_{pl} = M_4$ and $M_u \sim A^5 m_{pl} = M_5$. Symbols $M_4$ and $M_5$ are used from Eq(1) in previous section.

$M_{cr}$ is the total mass of the universe with length scale of $R_{cr} \sim r_{star}$ at time $t_{cr}$, when the density of the universe $\rho_{univ}$ is the critical density $\rho_{cr}$. If we suppose that $\frac{\rho_{univ}}{c^2} = const.$ for $\rho_{univ} > \rho_{cr}$, our universe will naturally have an inflation stag ($\dot{R}^2 \propto R^2$). Adopt $\rho_{univ} = \rho$, so “$\frac{\rho}{c^2} = const.$ for $\rho > \rho_{cr}$” means that there is a minimum radius $r_{\min}$ for all black holes. We have $r_{\min} \sim c/\sqrt{G\rho_{cr}}$, and the mass of lightest black hole (LBH) $M_{LBH} \sim r_{\min} \cdot c^2/G$. **Suppose** $r_{\min} = r_p$, then $M_{LBH} \sim r_p \cdot c^2/G = A \cdot m_{pl}$, and $\rho_{cr} \sim A^2 \cdot \rho_p$ in which $\rho_p \sim \frac{m_p}{r_p^3}$.

$M_u$ is the total mass of our universe, i.e. the mass of the "u particle"{1} with length scale of $R_u \sim r_p$ and mass scale of $M_u \sim A^5 m_{pl}$. Cosmic ring{2} may reflect the traces of $u$-particles collision. Thus, our universe may be a finite ball (balloon).

## A-particle

From Planck mass $m_{pl} = \sqrt{\frac{\hbar c}{G}}$, Planck length $r_{pl} = \sqrt{\frac{G\hbar}{c^3}}$, and $\frac{1}{2} m_{pl} \cdot c \cdot r_{pl} = \frac{1}{2}\hbar$, we can suppose that there is a Planck particle (Pl-particle) with mass $m_{pl}$, radius $r_{pl}$, and spin $\frac{1}{2}\hbar$. Since $r_p \sim \frac{\hbar}{m_p c}$, then $\frac{r_p}{r_{pl}} \sim A$. It means that a proton includes $A^3$ "Planck particles" with effective mass $m_A \sim \frac{m_p}{A^3} \sim 10^{-81} g$ each, which is the most elementary particle and can be named $A$-particle (it is a compound particle or string) with mass scale $m_A$, length scale $r_A \sim r_{pl}$, and compound spin $\frac{1}{2}\hbar$. A-particles control space-time (Ref {3}). According to Eq (1), the $m_A = m_8$ corresponds to an object with the maximum mass $M_{\max} \sim \frac{m_{pl}^3}{m_A^2} \sim A^8 m_{pl} = M_8$.

From the extension of mass scales sequence, we know our universe (a "$u$ particle") includes $A^3$ stars, a star includes $A^3$ protons, a proton includes $A^3$ $A$-particles. And an object with $M_{\max}$ includes $A^3$ $u$-particles.

## $\delta$-particles and Beyond SM

If we take $m' = m_e$, from Eq(1) we have that the corresponding celestial body is $M' \sim M_{starcluster}$. If we take $M'' \sim M_{galaxycluster} \sim 10^{15}$ solar mass, then the corresponding particle is $m'' \sim m_\delta \sim 10^0 - 10^1 eV$. We remember that $m_p$ corresponds to $M_{solar}$ and $m_\nu$ corresponds to $M_{\mathrm{super}structure}$. So, for stable particles $e, \nu, p, \delta,$ we have a "Twofold Standard Model Diagram"{1} in this multi-component universe. The "Twofold Standard Model Diagram" is as follows:

$$\begin{array}{cccc} u & c & t & \gamma \\ d & s & b & g \\ \nu_e & \nu_\mu & \nu_\tau & Z^0 \\ e & \mu & \tau & W^\pm \end{array} \qquad \begin{array}{cccc} u' & c' & t' & G \\ d' & s' & b' & g' \\ \delta & \delta' & \delta'' & Z' \\ p & p' & p'' & W' \end{array}$$

The speed of photon $\gamma$ is $c$. The speed of graviton $G$ is $c'$. We suppose the difference between the number of baryons $B$ and the number of lepton $L$ is conserved (B-L=const.). From this model we know: (1) there are new interactions in the right section, nuclear force is different from color force between quarks and directly connects to the gravitational interaction; (2) there is cosmic gravitational wave background (CGB) in the universe, and the GZK-limit{4} will be increased by two orders of magnitude; (3) There may exist $\pi_0', \pi_+', \pi_-'$ particles and heavy electron with mass $\sim 10^0 TeV$. (4) $\delta$-particle is like an inert neutrino but has a baryon number. $\delta$-particles can be a new energy source (Ref {5}).

## The diagram of mass scales sequence in the universe

Now we have extended the mass scales sequence in the universe as follows:

$m_{pl}$

(super-heavy particle) $A^{-0.5} m_{pl}$ $\qquad$ $A^{+1} m_{pl}$ (lightest black hole-LBH)

(proton) $A^{-1} m_{pl}$ $\qquad$ $A^{+2} m_{pl}$ (star)

(neutrino) $A^{-1.5} m_{pl}$ $\qquad$ $A^{+3} m_{pl}$ (superstructure)

$A^{-2} m_{pl}$ (super-light particle) $\qquad$ $A^{+4} m_{pl}$ (critical status, inflation)

$A^{-2.5} m_{pl}$ $\qquad$ $A^{+5} m_{pl}$ (u-particle)

• $\qquad$ •

• $\qquad$ •

$A^{-4} m_{pl}$ (A-particle) $\qquad$ $A^{+8} m_{pl}$

On the left side of the diagram are the mass scales of stable particles in micro-cosmos. On the right side are the corresponding mass scales of celestial bodies in macro-cosmos. A type of stable particle with mass scale $m_n = A^{-\frac{n}{2}} m_{pl}$ corresponds to a type of celestial bodies with mass scale $M_n = A^n m_{pl}$, n = 0, 1, 2,…7,8.

**Measuring $c$ and $\hbar$ year after year**

From the diagram of mass scales sequence in the universe, at n=4,5,6,7,8, $c$ has mutation and FTL. So we postulate that $c$ is always in evolving.

The universe is evolving and space-time (vacuum background) is also evolving. Thus, the fundamental physical constants ($\hbar, c...$) are naturally evolving. We have $\hbar c^{\alpha} = const.$ At the early epoch of the universe, $\alpha = 1$. After the universe transparency, $\alpha \to -3$.

The Planck length $r_{pl}$ may not represent the original scale of the universe. It reflects the time-space lattice scale $\lambda_0$ and the length scale $r_A$ of the most elementary particle: $r_{pl} \sim \lambda_0 \sim r_A$ (Ref {3}). At the early epoch of the universe (between $R_4$ and $R_5$) these scales were also evolving with cosmic scale $R$ ($\lambda_0 \propto R^{\beta}$). During this epoch, we have $\beta = 4$. After the universe became transparent, $\beta \to 0$. Thus, $R \propto c^{-\frac{3+\alpha}{2\beta}}$, $\frac{\Delta c}{c} = \xi \frac{\Delta R}{R}$, $\xi = -\frac{2\beta}{3+\alpha} < 0$; $\frac{\Delta \hbar}{\hbar} = -\alpha \frac{\Delta c}{c}$.

The evolution of $\hbar$ (since then the Rydberg constant) directly influences the value of cosmological redshift for all celestial bodies. Both of $c$ and $\hbar$ make contributions to the “abnormal” redshift $\Delta z$ of SN Ia, which has an approximate expression for small redshift $\Delta z \sim -\frac{\Delta c}{c}(z - 3\alpha)$. As an example, for $z \sim 1$ and $\Delta z \sim 0.2$, we have $\frac{\Delta c}{c} \sim 2\%$. In consideration of the influence of large redshift and the Stefan-Boltzmann constant, $\frac{\Delta c}{c}$ will be $\sim 1\%$.

Let the speed of light at the present time be $c_0$. The rate and the value of change for $c_0$ are $\frac{dc_0}{dt} \sim \xi \cdot H_0 \cdot c_0$ and $\Delta c_0 \sim \xi \cdot H_0 \cdot c_0 \cdot \Delta t$. We suggest measuring them year after year, and check whether $\frac{dc_0}{dt} < 0$ or not. Then, we can also check whether the cosmological constant/dark energy exists or not.

## Some Experiments and observations are suggested

1) To search for new particles in the desert area.
2) Year after year, accurately measuring the value of $c$ and $\hbar$.
3) Measuring the gravitational signal propagation speed $c'$, check GZK-limit.
4) Measuring the annihilation signals of dark matter particles ($\nu$, $\delta$).
5) Measuring cosmic ultra-high energy electrons and positrons related to cosmic $\delta$ particles.
6) Measuring the antineutrino and neutrino propagation speed difference.

## Reference (for extension section)

# Summary

## Mass scales sequence expresses the evolution of the universe

In our universe, the fundamental physical constants are the speed of light $c$, the gravitation constant $G$, and the Planck constant $\hbar$; the fundamental block of mass is the most stable baryon proton with mass $m_p$ and radius $r_p$ ($r_p \sim \frac{\hbar}{m_p c}$). From $c$, $G$, $\hbar$, we have

$$\text{Planck mass } m_{pl} \sim \sqrt{\frac{\hbar c}{G}} \sim 10^{19} GeV \qquad [1]$$

$$\text{Planck length } r_{pl} \sim \sqrt{\frac{\hbar G}{c^3}} \sim \frac{G m_{pl}}{c^2} \qquad [2]$$

then

$$\text{Large Number } A \sim \frac{m_{pl}}{m_p} \sim \frac{r_p}{r_{pl}} \sim 10^{19} \qquad [3]$$

The mass scales sequence of the universe was suggested more than twenty years ago[1], now it can have the diagram (mass tree) as follows[2]:

$$\begin{array}{ll}
m_0 = M_0 & \\
m_1 & M_1 \\
m_2 & M_2 \\
m_3 & M_3 \\
m_4 & M_4 \\
m_5 & M_5 \\
m_6 & M_6 \\
m_7 & M_7 \\
m_8 & M_8
\end{array}$$

On the left side of the diagram are the mass scales of stable particles in micro-cosmos:

$$m_n = A^{-\frac{n}{2}} m_{pl}, \text{ n = 0, 1, 2 ...} \qquad [4]$$

(with length scale $r_n$)

On the right side are the corresponding mass scales of celestial bodies in macro-cosmos:

$$M_n = A^n m_{pl}, \text{ n = 0, 1, 2, 3...} \qquad [5]$$

(with length scale $R_n$ and density $\rho_n$)

For n=0, $m_0 = M_0 = m_{pl}$ (export A- particle)

For n=1, $m_1 = A^{-0.5} m_{pl} \sim 10^{18} eV$ (critical energy[2],[3], super-heavy particle $m_1$)

$M_1 = A \cdot m_{pl}$ (lightest black hole - LBH[2])

For n=2, $m_2 = A^{-1} m_{pl} = m_p$ (proton)

$M_2 = A^2 m_{pl} = A^3 m_p = M_{star}$ (star)

For n=3, $m_3 = A^{-1.5} m_{pl} = A^{-0.5} m_p = m_\nu \sim 10^{-1} eV$ (neutrino)

$M_3 = A^3 m_{pl} = A \cdot M_{star}$ (superstructure[3]–[10], $M_3 = M_F$)

For n=4, $m_4 = A^{-2} m_{pl}$ (shrink/swell particle)

$M_4 = A^4 m_{pl} = A^3 M_1$ (end of inflation, $M_4 = M_{cr}$)

For n=5, $m_5 = A^{-2.5} m_{pl}$ (swell/shrink particle)

$M_5 = A^5 m_{pl} = A^3 \cdot M_{star} = M_u$ (the beginning of inflation, $M_5 = M_u$ [11],[2] )

.

.

.

For n=8, $m_8 = A^{-4} m_{pl} = A^{-3} m_p = m_A$(A-particle[12],[2], the most elementary particle)

$M_8 = A^8 m_{pl} = A^3 \cdot M_u$ (original universe, $M_8 = M_{\max}$)

From above diagram, there is a main mass sequence in the universe:
$m_8 \to m_2 \to M_2 \to M_5 \to M_8$
i.e. $m_A \times A^3 = m_p$; $m_p \times A^3 = M_{star}$; $M_{star} \times A^3 = M_u$; $M_u \times A^3 = M_8$.

It is obvious that the large number[12]–[15],[1] "$A$" plays an important role in both micro-cosmos and macro-cosmos as a fundamental physical constant. In cosmology, the fundamental physical constants are $G, c, \hbar, m_p$ or $G, c, \hbar, A$.

For n=0,1,2 these represent the fundamental blocks in the universe:
n=0, $M_0 = m_{pl}$, $R_0 = r_{pl} \sim \lambda_0$; $\lambda_0$ reflects the fundamental block of space-time[2],[12];
n=1, $M_1 = A \cdot m_{pl} = M_{LBH}$, is the fundamental block of the early universe[2];
n=2, $M_2 = A^2 m_{pl} = M_{star}$, is the fundamental block of the visible universe.

For n=4,5,6,7,8 that represent the evolution of the universe before H-decoupling; and from $R_5$ to $R_8$ all have a minimum radius $r_{\min}$ [3],[2] ($R_5 = R_6 = R_7 = R_8 = r_{\min}$).
$m_4$, $m_5$, $m_6$, $m_7$, $m_8$ are background particles.

For n=3, the Superstructures $M_3$ are the intermediate station in the evolution process of the universe. $M_3$-superstructures are slowly evolving structures and speed up the formation of the sub level celestial bodies [5] in it.

## The three stages of the evolution of the universe

The first stage is the chaos and volatilization stage.
In this stage (from n=8 to n=5), it is chaotic and volatilizing background particles $m_5, m_6$, $m_7, m_8$ into a big universe. Since $M_8 >> M_7 >> M_6 \ldots$, the main background particle is $m_8$ particle, i.e. the A-particle indeed;

The second stage is the inflation stage.
In this stage (from n=5 to n=4), we imagine an inverse process of evolution of the universe, when $\rho_{univ}$ arrive at $\rho_{cr}$ (n=4): $M_4 \sim A^4 m_{pl} \sim A^3 M_1$, $R_4 = R_{star} \sim A \cdot R_1$ and $\rho_4 \sim \frac{M_4}{R_4^3} = \frac{M_1}{R_1^3} \sim \rho_1 = \rho_{cr}$. Hence at this time, our universe includes $A^3$ lightest black holes. Since then, the LBHs were merged and collapsed into a "u particle", which has mass scale $M_u = M_5 \sim A^5 \cdot m_{pl}$ and length scale $R_5 \sim r_p$. It is obvious that the positive process of evolution from $M_5$ to $M_4$ (inflation process) is a step by step fission process of black holes (more and more LBH appear) and the CMB may have a fine grained structure;

The third stage is the expansion stage.
At n=4, LBHs break out and mix into a universe soup, which evolves gradually into the present universe. When the universe soup cools down gradually, the symmetries appeared as in a SM or BSM. At first in (arXiv: 1003.5208v3, July 2011) we have a Dual SM; then in (arXiv: 0804.2680v6, Aug 2011) we suggest a Two-Fold SM.

## Dark matter particles with low mass

The diagram of mass scales sequence is like a "mass tree". The diagram of different SM of particle physics is like "pods" (with symmetry) on the tree.
We suggest Dual Standard Model [11] diagram as follow:

$$\begin{matrix} u & c & t & \gamma \\ d & s & b & g \\ \nu_e & \nu_\mu & \nu_\tau & Z^0 \\ e & \mu & \tau & W^\pm \end{matrix} / H \qquad \begin{matrix} u_l & c_l & t_l & G \\ d_l & s_l & b_l & g_l \\ \delta & \delta' & \delta'' & Z' \\ p & p' & p'' & W' \end{matrix} / H'$$

Where $u_l$ $c_l$ $t_l$ $d_l$ $s_l$ $b_l$ are lept-quarks, $g_l$ is lept-gluon, $G$ is graviton. $Z', W'$ are the gauge bosons about a new type of interaction related to $\delta$ particles. The speed of photon $\gamma$ is $c$. The speed of graviton $G$ is $c'$. For left section and right section, there are Higgs bosons $H$ and $H'$ respectively. We also suggest Two-fold Standard Model [3]:

$$\begin{matrix} u & c & t & \gamma \\ d & s & b & g \\ \nu_e & \nu_\mu & \nu_\tau & Z^0 \\ e & \mu & \tau & W^\pm \end{matrix} / H \qquad \begin{matrix} u' & c' & t' & G \\ d' & s' & b' & g' \\ \delta & \delta' & \delta'' & Z' \\ p & p' & p'' & W' \end{matrix} / H'$$

In these models $\delta - particle$ and $\nu - particle$ are dark matter particles with low mass ($10^0 eV - 10^{-1} eV$)[11]. During the cooling process of LBH (and also of the collision fire ball in laboratory), if a lot of electrons are created before protons, the Dual SM is more supported.

## Search for $\delta$-particle

(1)According to Dual SM/two-fold SM, the $\delta$ particle is like an inert neutrino but with a baryon number. We may use high energy protons (>0.5 TeV) to collide carbon/beryllium targets to produce $\delta$ particles. Another option is to refer the equipment that was used to search for cosmic neutrinos.
(2)For the ultra-high energy primary cosmic ray spectrum (UEPCRS), the "knee" could be related to the interaction between proton and $C\nu B$: ( $p + \nu / \bar{\nu} \to \bar{e} + n / \bar{n}$ ) and the "ankle" could be related to the interaction between proton and $C\delta B$: ( $p + \delta / \bar{\delta} \to p + n / \bar{n}$ ). If space electron spectrum is correlated with UEPCRS, it will appear as two abnormalities by $10^0 TeV$ and $10^2 TeV$. The latter is related to $\delta$ particle. Besides, for cosmic ultra-high energy particles (such as $\nu_\mu$ [17]), the key is the production of ultra-high energy neutrons, which may be created in the interaction process of $\delta$ and $p$.
(3) If we deduce the influence of photons-baryons from CMB anisotropy spectrum, then it may reflect the mass spectrum of $\nu$ particles and $\delta$ particles.
(4)Based on the mass tree, it is preferable to look for $\delta$-particles in galaxy-clusters.

## Critical density $\rho_{cr}$ and critical energy $E_{cr}$

The energy scale at $R_4$ is the critical energy $E_{cr}$, and $\rho_{cr} \sim A^2 \cdot \rho_p \sim \rho_1$ in which $\rho_p \sim \frac{m_p}{r_p^3}$. From $\rho_{cr}$ we have $E_{cr} \sim (m_p^2 c^9 \hbar G^{-1})^{1/4} \sim 10^{18} eV$, which is like the cutoff of renormalization. At $E_{cr}$, when LBH are produced, the boundaries of elementary particles for SM or BSM in our spherical universe have disappeared. Without skin, where do the hairs adhere? So, This means that the interacted fields are also "disappeared" (unified). When the density at the center of a black hole became $\rho_{cr}$, the total mass and total gravitation of the black hole appear unchanged while accretion is processing. It is a type of pseudo-balance black holes/celestial bodies.

## Minimum black hole radius $r_{\min}$ and LBH

Proton is an elementary particle with complex structure but so stable, we adopt $r_{\min} = r_p$ means we live in a sphere universe, which belong to a big universe, mainly characterized by proton. One can adopt different $r_{\min}$ and that is corresponding to a different spherical universe and so on.

LBHs play a special effect in inflation process and at beginning of expansion process for cosmic evolution. Super heavy particle $m_1$ is a puzzle.

### Spherical universe

(1) GRBs Ring, multi-components universe and pancake.
Recently it was reported a giant ring-like structure with a diameter of 1.7 Gpc displayed by GRBs[16]. This giant ring can be explained by pancake process in a (B+$\nu$) two components superstructure[4], $m_\nu \sim 10^{-1} eV$, and our universe is inhomogeneous. We think that the cosmology principle needs to be reassessed.

(2) CMB cold spot and spherical universe.
The report about CMB cold spot have been around a long time[17],[18]. One of the possible explanations is that our universe is finite (spherical universe with present $R_{CMB} \sim 10^{28} cm$) and the earth is not at the center, but near the center. If that is the case, we can deduce the CMB anisotropy spectrums of cold semi-sphere is different to that of hot semi-sphere. When we make observation face or away from the CMB cold spot, the first peak of the CMB anisotropy spectrum will drop or rise while the sixth peak will rise or drop respectively. The CMB anisotropy spectrum itself is “anisotropic”.

(3) The civilization layer (ball).
We live in a spherical universe. According to the size of the CMB cold spot, recently the distance we deviate from the center of the sphere (using light year) is in numerical less than the H-decoupling time (using year).
The center region of the spherical universe is the civilization layer (ball). According to astronomical scale, it is just nearby the Earth.

(4) It is possible that spherical universe itself does not have dark energy, which is the superstructure effect. However, dark energy may associate with the background particles/field of the big universe in which our spherical universe is located and sometimes immersed in “$\Lambda$” background field or cloud.

(5) We may be able to obtain information from other spherical universe, especially its information in very early epoch

### Discussion

(1) The speed of light $c$ is not a constant. It has a mutation (phase transition) when $E = E_{cr}$ at the point $n = 4$ of mass tree. At point $n = 5,6,7,8$ of the mass tree, there could be different phase transitions.

(2) The speed of light represents the limit of the speed of movement of elementary particles. The maximum speed of particles that belong to different category or different cosmic level ( $n = 4,5,6,7,8$ ) are different.

(3) As for $\hbar c^{\alpha} = const.$, there are two region for $\hbar$ : at the early universe, when $c$ monotonously decreases, the $\hbar$ monotonously increases. After the universe became transparent, the $\hbar$ monotonously decreases as $c$ decreases and finally tends to zero.

(4) The microwave background (CMB) is on an atomic level. Correspondingly, the gravitational wave background (CGB) is on a nuclear level. As the collisions of electrons and positrons generate photons, the collisions of protons and anti-protons (or deuterons and anti-deuterons) create gravitons.

(5) The laws of nature can be expressed by mathematics under some simplifications. In other words, any physics theory is just an approximation. We discussed some phenomena related to super-high energy physics, which may determine the applicable region for modern physics including relativity.

## Reference (for summary section)